\documentclass[12pt,preprint]{aastex}

\usepackage{graphics}
\usepackage{psfig}

\slugcomment{draft version}

\shorttitle{G54.1+0.3}
\shortauthors{Lu et al.}

\begin{document}


\title{{\it Chandra} Observation of SNR G54.1+0.3 --- a Close Cousin of  the Crab Nebula}


\author{F.J. Lu\altaffilmark{1,2}, Q.D. Wang\altaffilmark{1}, B. Aschenbach\altaffilmark{3}, 
Ph. Durouchoux\altaffilmark{4}, and L.M. Song\altaffilmark{2}}


\altaffiltext{1}{Astronomy Department, University of Massachusetts,
    Amherst, MA 01003; lufj@flamingo.astro.umass.edu; wqd@gcs.astro.umass.edu}
\altaffiltext{2}{Laboratory of Cosmic Ray and High Energy Astrophysics, 
Institute of High Energy Physics, CAS, Beijing 100039, P.R. China; songlm@astrosv1.ihep.ac.cn}
\altaffiltext{3}{Max-Planck-Institut f\"ur Extraterrestrische Physik, Postfach 1312,
85741 Garching, Germany; bra@mpe.mpg.de}
\altaffiltext{4}{CE-Saclay, DSM, DAPNIA, Service d'Astrophysique, Gif-sur-Yvette Cedex, France;
durvla@discovery.saclay.cea.fr}

\begin{abstract}
We present a {\it Chandra} ACIS observation of SNR G54.1+0.3.
This supernova remnant is resolved into several distinct 
X-ray emitting components:
a central bright point-like source 
(CXOU J193030.13+185214.1), a surrounding ring, bipolar elongations, plus
low surface brightness diffuse emission. The spectra of these components are 
all well described by power-law models; the spectral index steepens 
with increasing distance from the 
point-like source. We find no evidence for any thermal plasma emission
that would correspond to shocked interstellar medium or ejecta. 
The observed morphological and spectral characteristics 
suggest that G54.1+0.3 is the closest ``cousin'' of the Crab Nebula --- 
a pulsar wind nebula driven by a combination
of equatorial and polar outflows from the putative pulsar represented by the 
point-like X-ray source.
 
\end{abstract}


\section{Introduction}
The Crab Nebula has played an essential role in our understanding of pulsar 
wind nebulae (PWNe). The basic observational X-ray characteristics  of 
the Crab Nebula are 
a central pulsar, an ellipse-shaped torus, two jets 
and the absence of an outer thermal shell. The X-ray torus is explained 
as shocked equatorial wind material from the central 
pulsar (e.g., Aschenbach \& Brinkmann 1975;  
Pelling et al. 1987; Weisskopf et al. 2000), 
while the jets correspond to the outflows from the two rotation poles of 
the pulsar (e.g., Aschenbach 1992). The observations and modeling of 
these characteristics have advanced our understanding of
the energetics and geometry of pulsar wind, as well as the 
physics of ultra-relativistic shocks and particle 
acceleration (e.g., Rees \& Gunn 1974; Aschenbach \& Brinkmann 1975; 
Kennel \& Coroniti 1984a, 1984b). 

However, morphologically the Crab Nebula is unique. No other source
has to date been found to mimic all the basic X-ray 
characteristics of the Crab Nebula (Gaensler 2001).
The PWNe surrounding the Vela Pulsar and PSR B1509-58 have 
X-ray arcs, yet these X-ray arcs are not complete X-ray rings
as observed in the Crab Nebula, and furthermore, these 
two PWNe are embedded inside bright thermal diffuse 
emission (Helfand, Gotthelf, \& Halpern 2001; Gaensler et al. 2001). 

In order to deepen our understanding of the properties of the 
pulsar wind and its
interaction with the environments, it is important to find a PWN 
which closely resembles the Crab Nebula. Supernova remnant 
(SNR) G54.1+0.3 is such 
a candidate. Radio emission from  G54.1+0.3 is centrally-filled, 
shows strong polarization, and has a flat spectrum, typical for a 
Crab-like SNR (Reich et al. 1985; Velusamy \& Becker 1988).  
A nonthermal X-ray spectrum from {\it ASCA} observations confirmed 
its Crab-like nature (Lu, Aschenbach, \& Song 2001).
As we will present in this letter, our recent {\it Chandra X-ray Observatory} 
observation of this remnant shows that it is the closest 
``cousin'' of the Crab Nebula. 

\section{Observation and Data Reduction}
SNR G54.1+0.3 was observed with  {\it Chandra} on 6-7 June 2001. 
The remnant was positioned at the aimpoint on the 
back-illuminated CCD chip S3 of the Advanced CCD Imaging 
Spectrometer (ACIS) in ``faint'' mode and at a 
working temperature of -120$\degr$ C. The spatial resolution is 
$\sim$0$\farcs$5, and the spectral 
resolution is $\sim$120 eV at 2.0 keV. The CCD frame read-out time for this 
observation is 3.2 seconds.

We calibrated the data using {\it CIAO} (2.1.3). After excluding 
two time intervals of high background we obtained a net exposure 
of 30.9 ks. Based on a comparison of the positions of 
the X-ray point source CXOU J193034.62+185110.8 (southeast 
of the remnant, see Fig. 1) and its optical counterpart, we find 
that the absolute positional accuracy of the X-ray observations 
is about 0$\farcs$6, consistent with the position accuracy 
claimed in the {\it Chandra Proposer's Observatory Guide}. 

\section{Analysis and Results}
While Fig. 1 presents an overview of the {\it Chandra} data,
Fig. 2 is a close-up of G54.1+0.3, with the image data adaptively smoothed
using a Gaussian to achieve a signal-to-noise ratio of $\gtrsim$5. 
The structure of the remnant is a complex, but we decompose it into 
several components (Fig. 1): a central bright point-like 
source (CXOU J193030.13+185214.1), 
a surrounding ring, elongations to the east and west, 
and extended low surface brightness emission in between and beyond the 
central structures. We find no evidence for a shell-like structure that 
would correspond to the thermal emission usually expected for an SNR. 
Therefore the remnant resembles the Crab Nebula in morphology 
quite closely. 

Fig. 3 displays spectral model fits of the individual components 
of G54.1+0.3 and Table 1 lists the corresponding model parameters. 
The background used in the spectral analysis is extracted from an annulus 
centered on CXOU J193030.13+185214.1 and with the inner and 
outer radii equal to 70$\arcsec$ and 130$\arcsec$ 
 (with CXOU J193034.62+185110.8 removed). 
 The spectrum of each of these components 
can be well fitted with a power-law model. There is no evidence 
for any emission line, and fitting the spectra with thermal models
 yields temperatures around 10 keV, 
which is much higher than the plasma temperature ($\lesssim$3 keV) typically
observed for young SNRs. Therefore we detect no significant thermal 
emission. We assume that the X-ray absorption column density is the same 
across the field as the absorption column densities
of all these components derived from individual spectral fitting are similar. 
 The jointly fitted value
 (1.6$\pm0.1\times10^{22}$ cm$^{-2}$)
 is about half of the total Galactic X-ray absorption 
($\sim$3.6$\times$10$^{22}$ cm$^{-2}$) derived from the {\it IRAS} 100 $\mu$m
emission in this direction (Boulanger \& Perault 1988; 
Wheelock et al. 1994). Assuming that the absorbing gas is 
relatively uniformly distributed along the line of sight, the
distance to G54.1+0.3 should be approximately 5 kpc, about half way
to the edge of the Galaxy from the Sun.

The source CXOU J193030.13+185214.1 most 
likely represents the pulsar that powers G54.1+0.3. The radial surface 
brightness profile of this source is consistent with the 
point spread function of the instrument. The slow CCD frame read-out 
time (3.2 s) of the observation, however, prevents a search 
for any periodicity typical of a young pulsar. We 
find no significant signal with a period $>$6 s. 
The count-rate of the source, 0.064$\pm$0.002 cts s$^{-1}$, indicates a 
pile-up fraction of only $\sim$6$\%$ with a negligible effect on the spectrum.  
The relation between the  
photon spectral indices of the source and the nebula 
is well consistent with the one derived by
 Gotthelf \& Olbert (2001) for pulsars 
and their respective nebulae. We find no emission peaks at the source position
at radio (with 4.8 GHz {\it VLA} data, Velusamy \& Becker 1988) 
and infrared (with the online {\it 2MASS} data)
wavelengths. From our 0.1-2.4 keV X-ray luminosity 
(1.8$\times$10$^{33}$ $d_{5}^{2}$ erg s$^{-1}$,where
$d_{5}$ is the distance in units of 5 kpc) of this point source, we infer  
the pulsar spin-down luminosity $\dot{E}$ as 
$\sim$2$\times$10$^{36}$ $d_{5}^{2}$ erg s$^{-1}$ 
(Becker \& Tr\"umper 1997). 

An eyeball fit to the ring with an ellipse centered on 
CXOU J193030.13+185214.1 gives the north-south semi-major axis 
as 5$\farcs$7 and the east-west semi-minor axis as  
3$\farcs$7. The eastern part of the ellipse is much brighter than the 
western part. The spectrum of the ring is the 
hardest among the diffuse components. 

The two elongations are not as well defined as the ring (Figs. 1-2). 
The western elongation is oriented perpendicular to the ring and extends 
$\sim$32$\arcsec$ horizontally from near the central point source to 
(RA, Dec; J2000) $19^{\rm h}30^{\rm m}28^{\rm s}$, 
$18\degr52^{\prime}14\arcsec$.  
The eastern elongation extends $\sim$14$\arcsec$ ending 
at about 19$^{\rm h}30^{\rm m}31^{\rm s}$, 
$18\degr52^{\prime}10^{\prime\prime}$. The orientation has   
 an average inclination of 
 $~$$\sim$18$\degr$ to the south. 

The outer low surface brightness diffuse emission marks the
accumulated pulsar wind material. The size of this low surface
brightness emission (1$\farcm5\times1\farcm2$) is much larger than 
that of the central bright region (0$\farcm7\times0\farcm3$) 
composed of the point-like source, the ring and the elongations.  
The low surface brightness component shows some extensions to the west, 
north and northeast and compares well with the radio map both in size 
and overall shape (Velusamy \& Becker 1988). The spectrum is 
significantly softer than those of the ring and western elongation, 
suggesting that, if we assume synchrotron radiation, the particles 
in this region are more evolved. A detailed comparison and 
explanation of the X-ray and radio maps and a more extensive 
study of the low surface brightness diffuse emission  
will be presented in a separate paper.

\section{Comparison with the Crab Nebula}
The X-ray morphology of G54.1+0.3 is strikingly similar to that of the
Crab Nebula although some differences exist.
Both remnants show a complete X-ray ring (torus) around the 
corresponding pulsar, plus two opposite elongations (jets), as well 
as the lack of a thermal 
outer shell. But the elongations of G54.1+0.3  are more diffuse 
than the Crab jets, that we will discuss more specifically 
in $\S$5.2. G54.1+0.3 shows large scale low surface 
brightness X-ray emission beyond the ring and the elongations.  
This emission coincides well with the radio morphology of the 
remnant (Velusamy \& Becker 1988). In contrast, little 
X-ray emission is presented beyond the Crab 
torus (Weisskopf et al. 2000).  The radio 
size of the Crab Nebula is $\sim$3 times bigger than 
the X-ray size (Bietenholz \& Kronberg 1990; Weisskopf et al. 2000). The
morphological difference of the outer low surface brightness
emission suggests that the synchrotron cooling efficiency 
in the G54.1+0.3 ring is much lower than in the Crab torus.
     
G54.1+0.3 and the Crab Nebula have quite similar spectral characteristics.
The spectra of both remnants are characterized by power-laws which
steepen progressively from pulsar, through torus (ring) 
and jets (elongations), to the outer diffuse emission 
(Table 1; Willingale et al. 2001).  The similar spectral steepening
trend indicates that particles follow 
similar tracks in the two remnants. 
However, the spectra of G54.1+0.3 are flatter than those of the corresponding 
components in the Crab Nebula, consistent with lower efficiency of the 
synchrotron cooling in G54.1+0.3. 

Both the total X-ray luminosity and the luminosities of 
the corresponding components in G54.1+0.3 differ substantially 
from those of the Crab Nebula (see Table 1 and Weisskopf et al. 2000).
The putative pulsar and the X-ray nebula of G54.1+0.3 are 
about two orders of magnitude less 
luminous than those of the Crab. The ring 
of G54.1+0.3 accounts for only about 10$\%$ of the total 
extended emission, but the 
torus of the Crab Nebula dominates the X-ray emission 
of the whole nebula. In G54.1+0.3, the luminosities of the 
elongations are comparable with 
that of the ring, whereas in the Crab Nebula, the jets 
are obviously much weaker than the torus. 
 These differences are likely due to the particular pulsar wind 
geometry and energy input to the various components in the two remnants.

\section{Pulsar Wind Properties}
 We suggest that the X-ray 
ring and elongations are due to the equatorial wind
and the two polar outflows from the pulsar, respectively,  
 as illustrated schematically in Fig. 4.   
 
\subsection{The X-ray Ring}
The elliptical ring in G54.1+0.3 is apparently the projection 
of an inclined circular ring. Similar to the X-ray torus 
in the Crab Nebula, this ring might be due to
the shocked pulsar equatorial wind (Aschenbach \& Brinkmann 1987). 
The ratio of observed semi-major and semi-minor axes 
suggests an inclination angle $\theta$=41$\degr$.

Following Pelling et al. (1987), we calculate whether
the Doppler boosting of the downstream motion can explain 
the X-ray brightness variation 
across the ring (Fig. 5).  The intensity ($I$) of the ring 
at position angle $\phi$ measured counter-clockwise 
from the north can be expressed as
\begin{equation}
 I=I_{0}*[\frac{\sqrt{1-v^2}}{1-{\rm cos}(\theta)*v*{\rm sin}(\phi)}]^{\alpha+1}+I_{b},
\end{equation}
where $I_0$ is the unshifted intensity, 
$v$ is the bulk motion velocity in units of $c$ --- the 
speed of light, $\alpha$ is the photon spectral index, 
and $I_{b}$ is the intensity of the diffuse background emission.
Fig. 5 shows that the Doppler boosting model fits  the measured intensity 
variation well. The fitted parameters and their
1$\sigma$ uncertainties are: $I_{0}$=7.6$\pm$2.4 cts arcsec$^{-2}$, 
$v$=0.40$\pm$0.12 c, and $I_b$=5.7$\pm$2.5 cts arcsec$^{-2}$. 
Using the $v$ value and following Kennel \& Coroniti (1984b) we infer
a ratio of the electromagnetic energy flux to the particle energy flux 
in the wind of $\sigma$$\simeq$0.06.
Therefore, the surface brightness variation across the ring 
is consistent with the assumption that the enhanced 
X-ray emission arises from the freshly shocked pulsar wind material.

\subsection{The Elongations}
We propose that the elongations are associated with the pulsar
polar outflows. This association is suggested by the morphology  
of the western elongation (Figs. 1-2). 
The spectrum of the elongation is flatter than that of the 
surrounding low surface brightness emission, suggesting that the 
particles in the western elongation have undergone fewer 
synchrotron losses. The 
eastern elongation is not well resolved, but appears bent 
and shorter than the western elongation. These two elongations, 
however, are broader and more irregular
than the relatively well confined jets observed in Crab (Weisskopf et al. 2000), 
Vela (Helfand et al. 2001), and PSR B1509-58 (Gaensler et al. 2001),
indicating that the 
pulsar polar flows  in G54.1+0.3 are less 
collimated. 

\subsection{Geometry} 
The X-ray morphology of the ring and the two elongations 
in G54.1+0.3 suggest that
the pulsar wind is concentrated in the equatorial
plane and the polar outflow directions. The width of the ring is 
about 2$\farcs$5 in the minor axis direction. As the radius of the 
deprojected ring is 5$\farcs$7, the width indicates that the equatorial
wind particles fan out in a half opening angle of $\lesssim$16$\degr$, 
corresponding to $\lesssim$1.1$\pi$ steradian. 
The width of the head of the western elongation 
is measured to 6$^{\prime\prime}$, 
implying that the half opening angle of the western polar outflow 
is $\sim$4$\degr$, assuming that the outflow is perpendicular to the 
equatorial plane as implied by Fig. 2. 
The corresponding solid angle is 5$\times10^{-3}$$\pi$ steradian. 
The open angle of the eastern outflow remains 
unclear because of the insufficient 
knowledge about its direction and thus length.

The observed pulsar wind geometry reflects the
fractional wind power injected to the ring and elongations.  
 Because the reverse shock occurs when the wind ram pressure balances
the diffuse nebula pressure, $p_n$, we have
\begin{equation}
r_{s}=(\frac{\delta \dot{E}}{\Omega c p_n})^{1/2},
\end{equation}
where $r_s$ is the distance from the pulsar,
 $\delta$ is the fraction of the wind power $\dot{E}$ in one particular
wind component, and $\Omega$ is the solid open angle of that component.
We assume that $\dot{E}$ is split between  
$\dot{E}_e$ (injected to the equatorial ring) and 
$\dot{E}_p$ (released in the polar outflows) that is 
equally divided between the two polar outflows. We also assume that $p_n$ 
does not change substantially with position in the diffuse nebula, 
because the sound speed in a relativistic plasma is $c/\sqrt{3}$ and so no pressure 
gradient in the bubble can be maintained on time scales that are dynamically 
important (e.g., Rees \& Gunn 1974). From Equation (2), 
with the geometric parameters and assumptions, we infer  
that $\gtrsim1/3$ of $\dot{E}$ is released to the polar outflows.

\section{Summary and Conclusions}
Based on a {\it Chandra} ACIS-S observation, we have resolved the spatial 
and spectral structures of SNR G54.1+0.3. We find that both its morphology 
and the spectra closely resemble those of the Crab Nebula, though the
synchrotron cooling efficiency in G54.1+0.3 may be substantially lower
than in the Crab Nebula. In particular, 
we have discovered a point-like source, CXOU J193030.13+185214.1, at a location
close to the morphological center of  G54.1+0.3. The 
source is most likely the pulsar powering the SNR, although a spin 
period of the pulsar is yet to be determined. 
We estimate the pulsar spin-down luminosity to be 
$\sim$2$\times$10$^{36}$ $d_{5}^{2}$ erg s$^{-1}$ from the 
X-ray luminosity of the point source. 
We suggest that the diffuse X-ray emission around the pulsar originates from 
the shocked pulsar wind. The presence of the X-ray ring indicates 
a strong equatorial wind with an inclination angle of 41$\degr$. From 
the surface brightness variation across the ring we infer the downstream 
flow speed of the shocked pulsar wind material to be  0.40$\pm$0.12 {\it c}, which indicates
that the wind is particle dominated. The two elongations are oriented 
nearly perpendicular to the ring and are likely to be associated with the 
polar outflows from the pulsar.  The geometry of the X-ray ring and 
the elongations suggest at least one-third of the  
spin-down energy is injected into the two polar outflows.
These characteristics of SNR G54.1+0.3 provide new constraints for modeling 
of the elusive pulsar wind  and its interaction with the environment.

\acknowledgments
 We thank the referee D. J. Helfand for his comments. This work is supported 
partially by the NASA-grant SAO GO-12068X and NASA LTSA grant NAG5-7935.  
FJL and LMS also appreciate support of the Special Funds for Major State 
Basic Research Projects and the National Natural Science
Foundation of China.

\clearpage


\begin{deluxetable}{lccc}
\tabletypesize{\footnotesize}
\tablecaption{Spectral properties of various components of G54.1+0.3\tablenotemark{a} \label{tbl-1}}
\tablewidth{0pt}
\tablehead{
\colhead{Components} &\colhead{Photon Index}  & \colhead{$F_{1}$\tablenotemark{b}} &
\colhead{$F_{2}$\tablenotemark{b}}}
\startdata
 Pulsar & 1.09$^{+0.08}_{-0.09}$  & 1.72 &2.31\\
 Ring&1.64$^{+0.18}_{-0.16}$&0.52&0.97\\
 Western Elongation&1.66$^{+0.16}_{-0.14}$&0.68&1.30\\
 Eastern Elongation&1.95$^{+0.28}_{-0.26}$&0.20&0.49\\
 Outer Region&1.97$^{+0.11}_{-0.12}$&2.31&6.36\\

\enddata


\tablenotetext{a}{Parameters are from fit with $\chi^2$=366.4 
and 386 degree of freedom. The jointly-fitted X-ray-absorbing 
column density is 1.6$\pm$0.1$\times10^{22}$ cm$^{-2}$.}
\tablenotetext{b}{$F_1$ and $F_2$ are absorbed and unabsorbed fluxes, respectively;
both in the 0.2-10.0 keV band and in units of 10$^{-12}$ ergs cm$^{-2}$ s$^{-1}$. }

\end{deluxetable}
\clearpage

\begin{figure}
\plotone{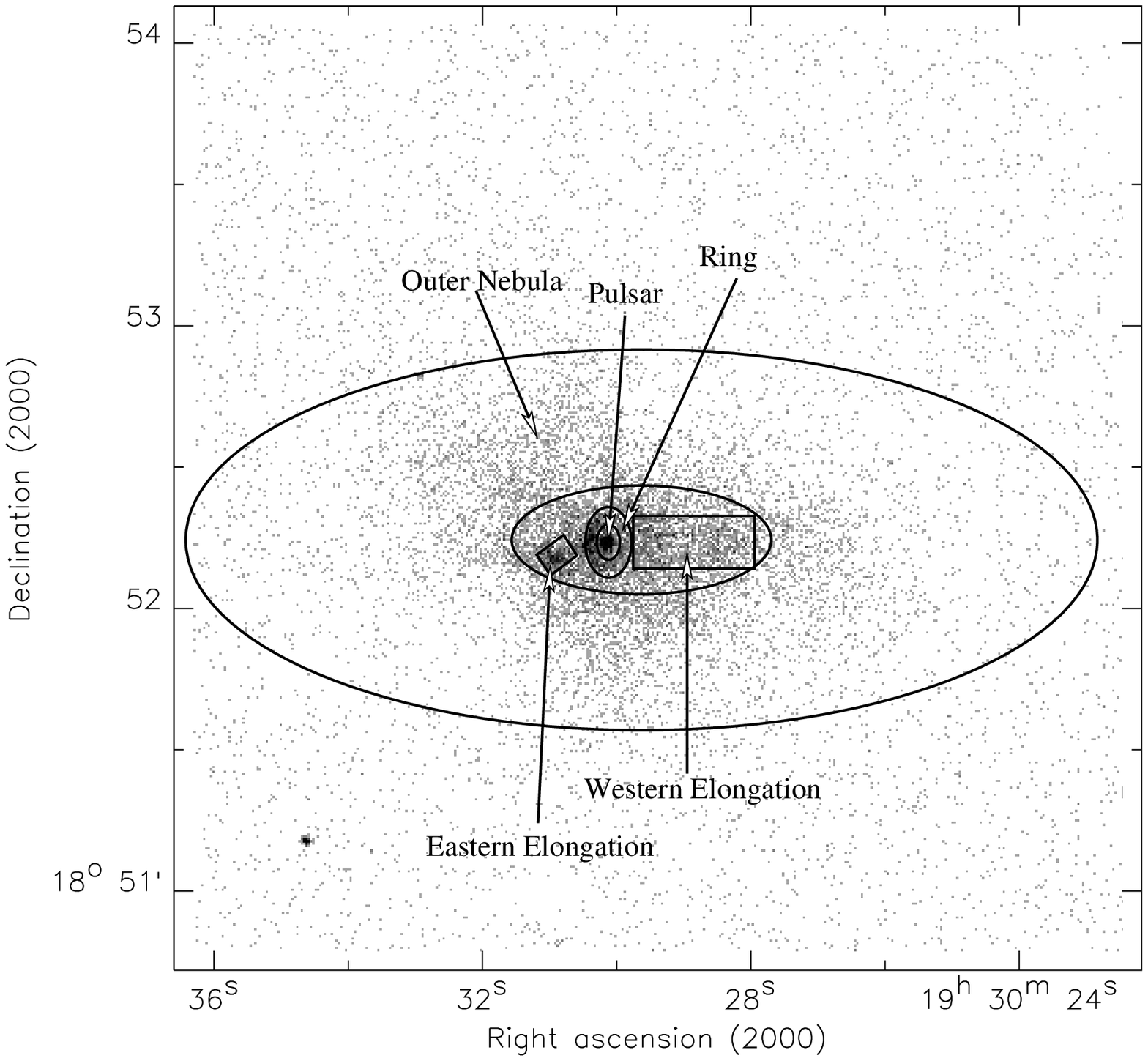}
\caption{{\it Chandra} ACIS-S count distribution in the field of SNR G54.1+0.3. 
The overlaid boxes and annuli define the regions from which spectral data are 
extracted (Table 1). 
\label{fig1}}
\end{figure}
\clearpage

\begin{figure}
\plotone{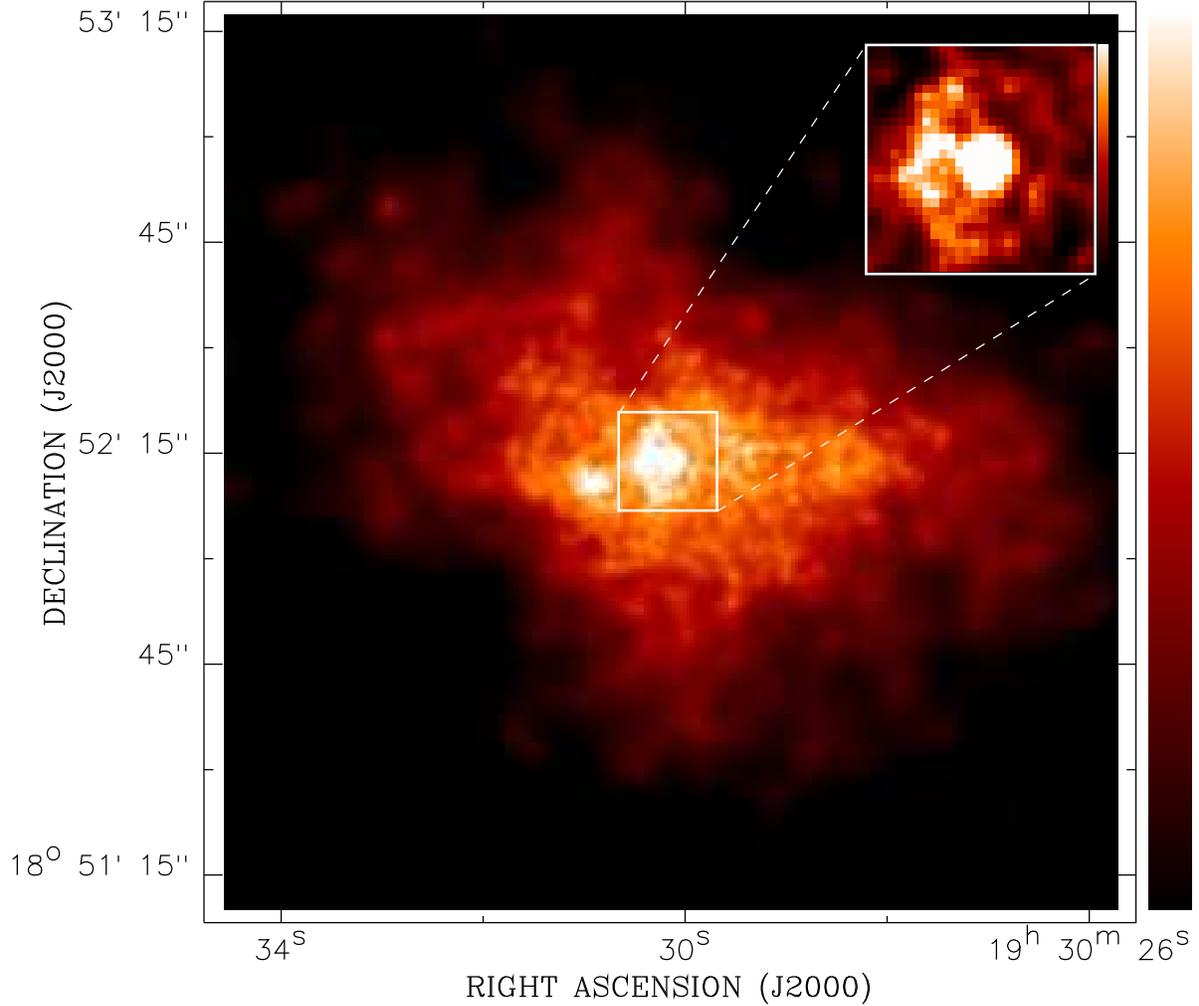}
\caption{{\it Chandra} ACIS-S X-ray (0.5-10 keV) intensity 
map of SNR G54.1+0.3. 
The image is smoothed adaptively with a Gaussian adjusted to achieve 
signal-to-noise ratio $\gtrsim$5. The intensity is plotted 
on logarithmic scale from 7.43$\times$10$^{-5}$ to 5.95$\times$10$^{-3}$ 
cts cm$^{-2}$ s$^{-1}$ arcmin$^{-2}$. Shown in the upper-right corner is the central region of 
SNR G54.1+0.3 plotted on logarithmic scale from 1.49$\times$10$^{-3}$ to 8.92$\times$10$^{-3}$ 
cts cm$^{-2}$ s$^{-1}$ arcmin$^{-2}$. \label{fig2}}
\end{figure}

\begin{figure}
\plotone{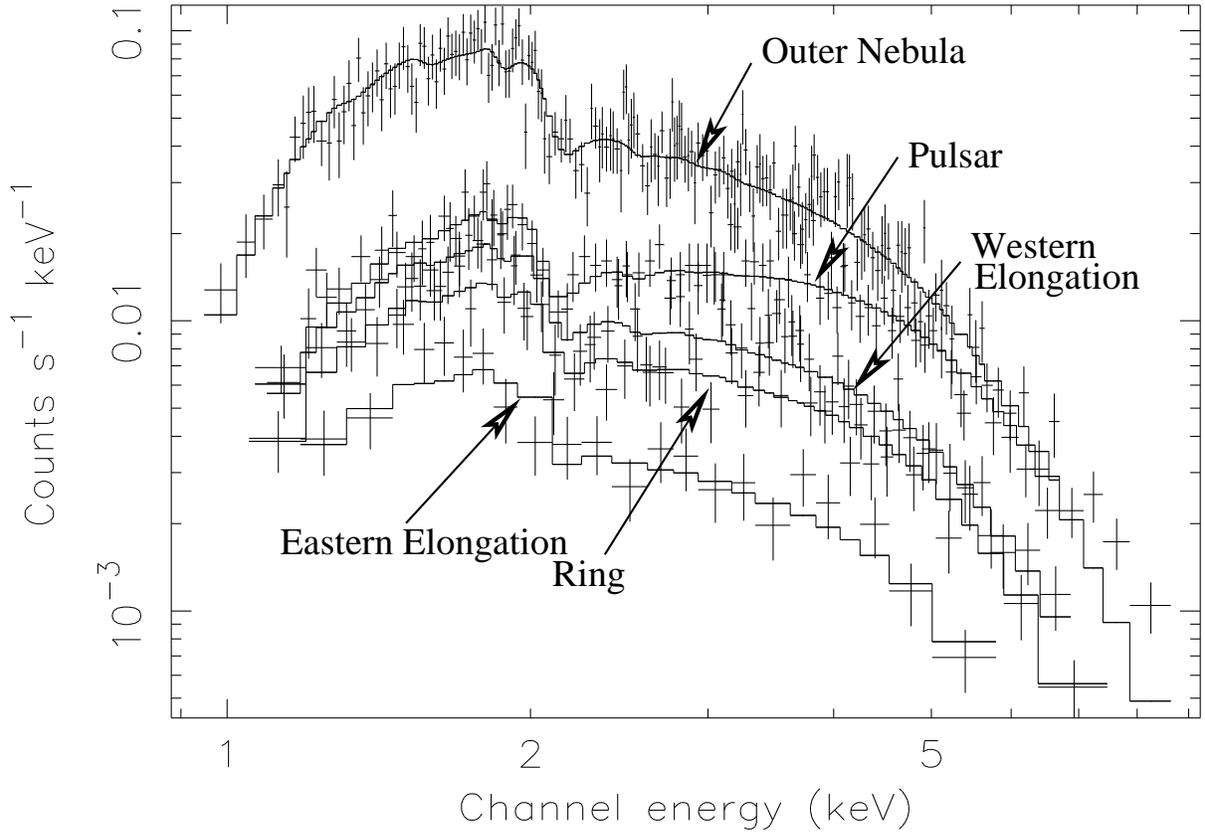}
\caption{Power-law model fits to the spectra of various 
components of SNR G54.1+0.3.  \label{fig3}}
\end{figure}

\begin{figure}
\plotone{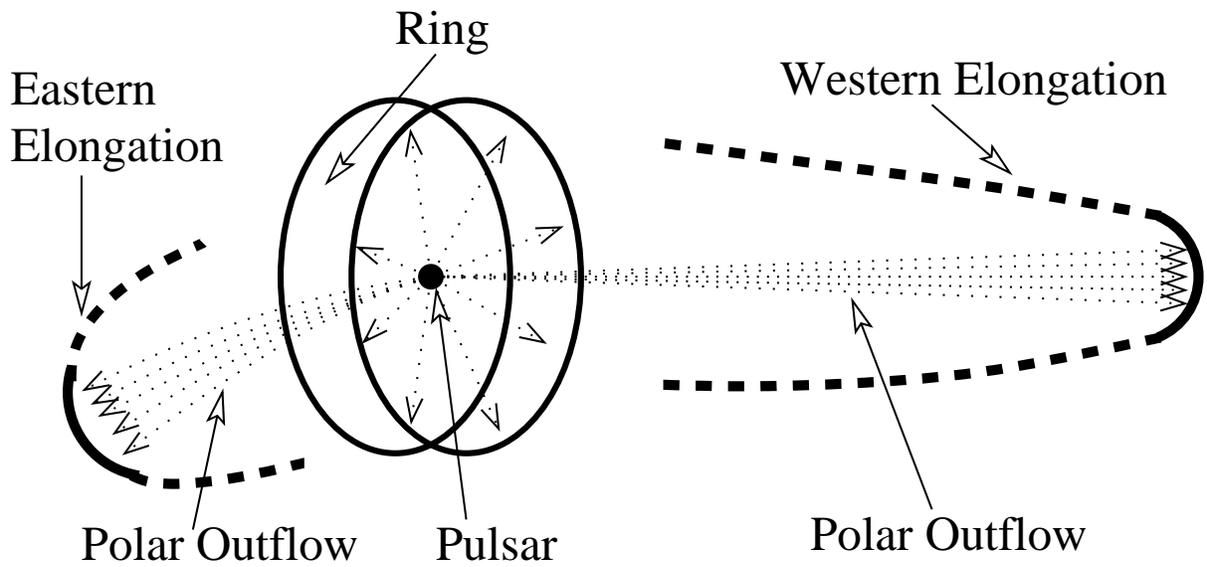}
\caption{Illustration of various components in 
the central region of SNR G54.1+0.3. 
The wind from the central pulsar (dotted thin lines)
 consists of two primary components: an equatorial wind
 and two polar outflows. The solid lines represent the 
terminal shocks and the dashed thick lines mark the
 regions where instabilities and/or 
bulk material drifted from the terminal shock giving rise
 to some bright emission features.
\label{fig4}}
\end{figure}
\clearpage

\begin{figure}
\plotone{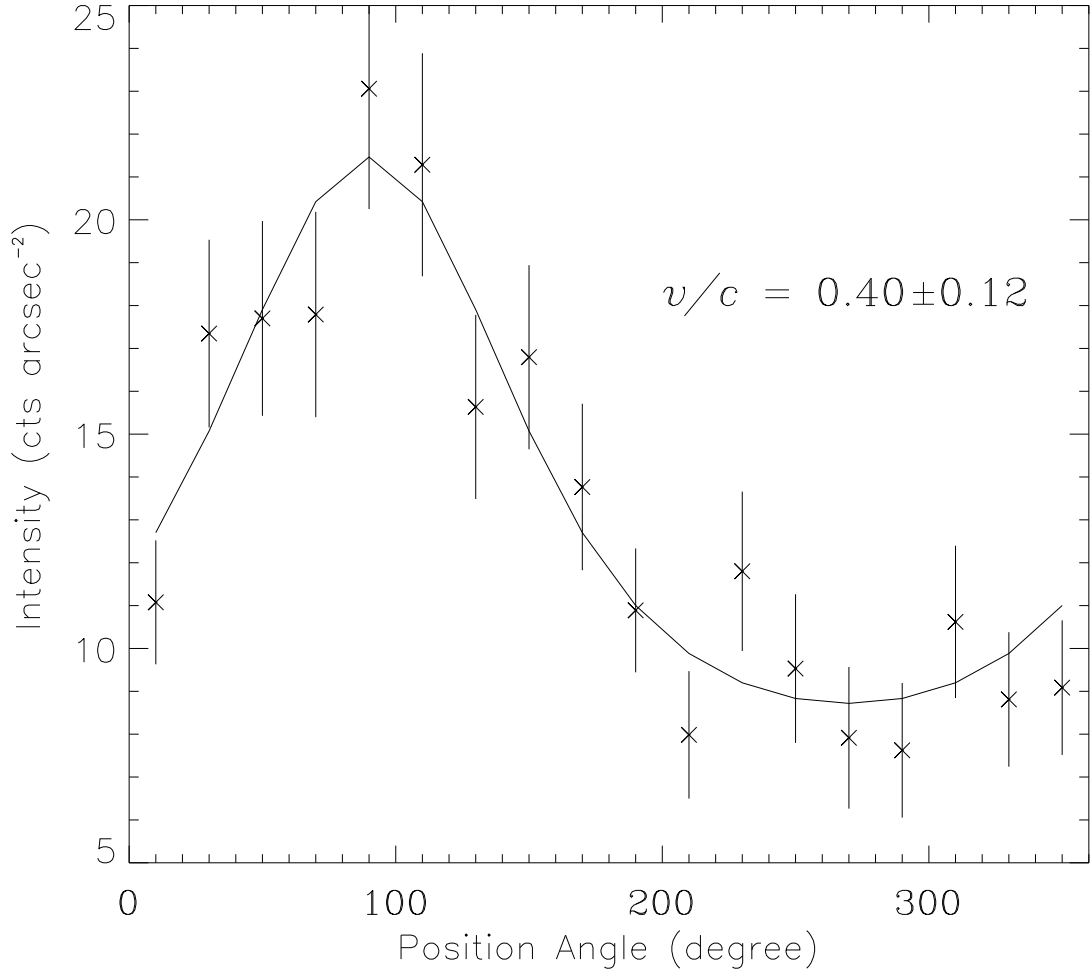}
\caption{The brightness distribution of the X-ray ring vs. 
position angle (from north to east). 
Error bars are at 1 $\sigma$ confidence level. The solid line is the best
fit to the data with a Doppler boosting model.                         
\label{fig5}}
\end{figure}

\end{document}